\documentstyle[twoside,fleqn,espcrc2,rotate,epsf]{article}

\newcommand{\AmS}{{\protect\the\textfont2
  A\kern-.1667em\lower.5ex\hbox{M}\kern-.125emS}}

\def\lsim{\raise0.3ex\hbox{$<$\kern-0.75em\raise-1.1ex\hbox{$\sim$}}}
\def\gsim{\raise0.3ex\hbox{$>$\kern-0.75em\raise-1.1ex\hbox{$\sim$}}}

\title
{
Status of the CP-PACS Project
\thanks{
{Presented 
at {\it Lattice 96}, St.\ Louis, USA.
}}
}

\author{
Y.~Iwasaki \address{Center for Computational Physics and
Institute of Physics, University of Tsukuba, Ibaraki 305, Japan} 
\ \ for the CP-PACS Collaboration
}

\begin{document}

\begin{abstract}
The CP-PACS computer with a peak speed of 300 Gflops was completed
in March 1996 and has started to operate. We describe 
the final specification and the hardware implementation of 
the CP-PACS computer, and its performance for QCD codes.
A plan of the grade-up of the computer scheduled for fall of 1996
is also given.
\end{abstract}

\maketitle

\section{CP-PACS project}
The CP-PACS project\cite{cppacs-pro} is a five-year project which 
formally started in 1992.  The project currently consists of 33 
members in physics and computer science as listed in 
Ref.~\cite{cppacs-mem}.
We selected Hitachi Ltd. as the industrial parter through a formal 
bidding process soon after the start of the project, and we 
have been working in a close collaboration for the hardware and software 
development of the CP-PACS computer.
The fundamental design of the computer was laid down 
in 1992, its details worked out in 1993, and the logical design and 
the physical packaging design was completed in 1994.
Chip fabrication and assembling of parts started in early 1995, and 
the CP-PACS computer with a peak speed of 300 Gflops was completed in 
March 1996.

\begin{table}[hbt]
\setlength{\tabcolsep}{0.1pc}
\caption{Specification of the CP-PACS computer}
\label{specification}
\vspace{-3mm}
\begin{center}
\begin{tabular}{l l}
\hline
peak speed &300Gflops(64 bit data)\\
main memory & 64GB \\
parallel architecture &MIMD with\\
         & distributed memory\\
number of nodes&1024 \\
node processor&HP PA-RISC1.1+PVP-SW\\
\hspace*{5mm}\#FP registers&128\\
\hspace*{5mm}clock cycle & 150MHz\\
\hspace*{5mm}1st level cache&16KB(I)+16KB(D)\\
\hspace*{5mm}2nd level cache&512KB(I)+512KB(D)\\
network&3-d hyper-crossbar\\
\hspace*{5mm}node array &$8\times 17\times 8^*$\\
\hspace*{5mm}through-put&300MB/sec\\
\hspace*{5mm}latency&$3\, \mu$sec\\
distributed disks&3.5'' RAID-5 disk\\
\hspace*{5mm}total capacity&529GB\\
software&\\
\hspace*{5mm}OS&UNIX micro kernel\\
\hspace*{5mm}language&FORTRAN, C, assembler\\
front end& main frame\\
& connected by HIPPI\\
\hline
\multicolumn{2}{r}{${}^*$ including nodes for I/O \ \ \ }
\end{tabular}
\end{center}
\vspace{-10mm}
\end{table}

\section{Hardware implementation}
 
\begin{figure}[t]
\begin{center}
\leavevmode
\epsfxsize=6.0cm
  \epsfbox{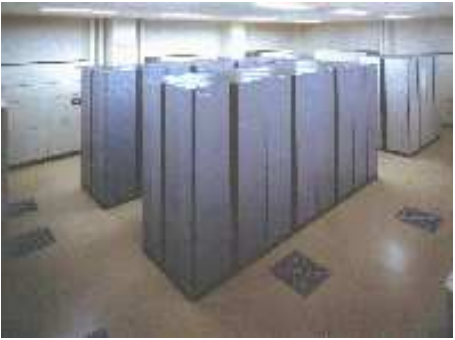}
\end{center}
\vspace{-0.5 cm}
\caption{Outlook of the CP-PACS computer}
\label{outlook}
\vspace{-0.4cm}
\end{figure}

\begin{figure}[t]
\begin{center}
\leavevmode
\epsfxsize=5.5cm
  \epsfbox{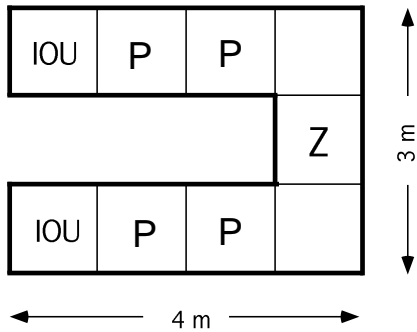}
\end{center}
\vspace{-0.5cm}
\caption{Floor-plan of the CP-PACS computer}
\label{cppacs-floor}
\vspace{-0.4cm}
\end{figure}

A picture of CP-PACS computer is shown in Fig.~\ref{outlook}.
The size of the computer is roughly
2m$\times$4m$\times$3m in height, width and depth. 
The floor-plan is depicted in Fig.~\ref{cppacs-floor}.
For the major architectural characteristics of the computer I refer to 
Ref.~\cite{cppacs-pro}.

The final specification of the CP-PACS computer
is summarized in Table~\ref{specification}.
The size of the second-level cache has been doubled since last year.
The number for latency of data transfer represents the measured value
in the remote DMA mode, 
which is the fastest mode for data transfer,  
including software and hardware overheads and averaged over 
transfer through $x$, $y$ and $z$ crossbar switches.

Figure \ref{floor-plan}
shows the floor plan of the CPU chip
which is fabricated by 0.3 micron CMOS semiconductor technology,
with the size being 1.57cm$\times$1.57cm. 
The PVP-SW feature,  which enables vector calculations very effectively 
within the RISC architecture of CPU, is implemented with 128 floating-point
registers in the green part at the lower right corner 
of Fig.~\ref{floor-plan}.

\begin{figure}[t]
\begin{center}
\leavevmode
\epsfxsize=6.0cm
  \epsfbox{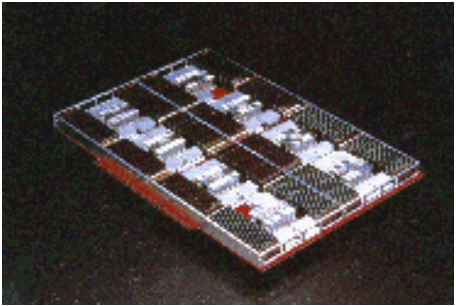}
\end{center}
\vspace{-0.5cm}
\caption{Silicon multichip module of CPU}
\label{module}
\vspace{-0.4cm}
\end{figure}

A silicon multichip module is depicted in Fig.~\ref{module}
where the chip located at the center is the CPU chip.
The adjacent two chips are the network interface adapter (NIA)
and the storage controller (SC)
which are fabricated by 0.5 micron gate-array technique.
Twelve chips surrounding them are off-chip second-level cache made of SRAM.
The size of the module is about 5.7 cm $\times$ 7.2 cm.

\begin{figure}[t]
\begin{center}
\leavevmode
\epsfxsize=6.0cm
  \epsfbox{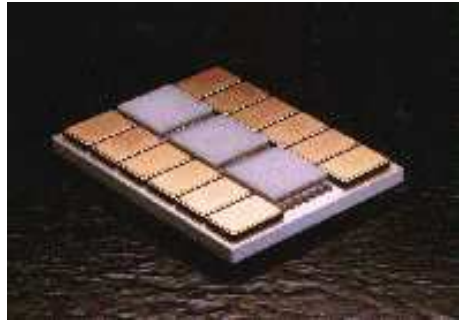}
\end{center}
\vspace{-0.5cm}
\caption{One board consists of eight CPU units}
\label{board}
\vspace{-0.4cm}
\end{figure}

One board which consists of 8 nodes is shown in Fig.~\ref{board}.
The center of the white part of each unit 
corresponds to the multichip module
shown in Fig.~\ref{module}, now with fins for 
air-cooling. The black part is the main memory with 4 Mbit DRAM.
The other three white chips on each unit are main-memory
address/data controllers. In addition each board has two chips for crossbar 
switches in the $x$ direction 
and one chip for clock distributer.
The size of one board is 45.6 cm $\times$ 62.5 cm.
Sixteen boards are installed in a crate and two crates are installed 
in a cabinet
represented by a square with symbol P in Fig.~\ref{cppacs-floor}.

\newbox\B
\epsfysize=5.2cm
\setbox\B=\vbox{\epsfbox{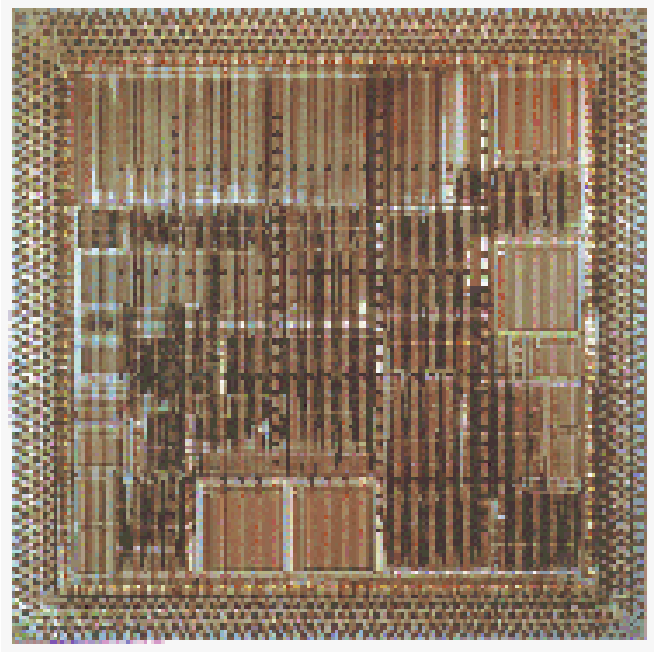}}

\begin{figure}[t]
\begin{center}
\leavevmode
\epsfysize=6.0cm
  \makeatletter
  \@rotu\B
  \makeatother
\end{center}
\vspace{-0.5cm}
\caption{Floor-plan of the CPU chip}
\label{floor-plan}
\vspace{-0.4cm}
\end{figure}

The crossbar switches in the $x$ direction are mounted on each board 
connecting 8 nodes,
as explained above,
those in the $y$ direction placed on a back-plane
located in four cabinets (symbol P in Fig.~\ref{cppacs-floor})
and
those in the $z$ direction mounted on a board which
is housed in one cabinet (symbol Z in 
Fig.~\ref{cppacs-floor}). 

In the two cabinets with symbol IOU, adaptors for I/O of data to the
distributed disks are installed. Raid-5 disks which are connected by
SCSI-2 bus through the adaptors are set in
cabinets installed a few meters apart.

\section{Performance}

We write codes for lattice QCD with Fortran 90 which includes
libraries for data communication.
A Fortran compiler 
incorporating the PVP-SW feature has been newly developed,
which produces efficient object codes.
The performance of the object code is typically 
90 -- 150 Mflops per node, depending on the structure of the do-loop.
The through-put of the data transfer between nodes
with Fortran libraries, in the case of data of 576 Kbytes as an example,
is 250 Mbytes/sec,
which is to be compared with the peak through-put of 300 Mbytes/sec.

The update time per link with a pseudo-heat bath method
for one processor is 55.3 $\mu$sec,
which corresponds to 103 Mflops/PU.
The performance for the Wilson quark matrix multiplication
for the red-black algorithm is
96 Mflops/PU with the present code.
On the other hand, we have a hand-optimized assembler code
for the Wilson quark matrix multiplication with which
the performance reaches 195 Mflops/PU, which is 65 \%
of the peak speed.
We are now modifying this assembler code for the red-black algorithm.
For MR red/black solver, the performance of the calculation part
is 122 Mflops/PU, which reduces to 93 Mflops/PU when
the data communication is included.
The percentage of communication is about 20 \% of the total time.
In the case of the CG solver for KS fermion, the performance is 128 Mflops
for the case when the length of the inner most loop is 128.

After checking the fundamental performance,
we have performed a test of the computer as a whole with 
a quenched QCD spectrum calculation
with the Wilson quark action on a $64^4$ lattice 
at $\beta =6.0$ for three hopping parameters
($m_\pi/m_\rho \simeq 0.7$, 0.5, and 0.4),
for two of which there exist already previous mass spectrum calculations.
Results for the effective masses of hadrons 
for the smaller two hopping parameters are in good agreement
with the previous results.
This makes us confident that the machine is working properly and that 
our codes are correct.

\section{Grade-Up of the CP-PACS computer}
We plan to grade-up the CP-PACS to 
a peak speed of 600 Gflops and a memory size of 128 Gbytes,
increasing the number of nodes from 1024 to 2048 in the coming fall.
The total funding approved 
including that for the grade-up is 2.2 billion yen
(about 22 million US dollars).
Until the grade-up  we plan to run a quenched spectroscopy calculation 
with Wilson quarks at four values of $\beta$ in the range of 
$m_\pi/m_\rho =0.4$ to 0.75 on lattices with a spatial size 3.0 fm.

This work is supported by 
the Grant-in-Aid of Ministry of Education,
Science and Culture (No.07NP0401).


\begin{thebibliography}{99}
\bibitem{cppacs-pro}Y. Iwasaki, Nucl. Phy. B(Pro. Suppl)34(1994) 78,
A. Ukawa, {\it ibid.} 42(1995)194.
\bibitem{cppacs-mem} The present members of the CP-PACS project are:
S. Aoki, T. Boku, M. Fukugita, S. Gunji,
T. Hoshino, S. Ichii,
M. Imada, S. Ishizuka,
Y. Iwasaki, K. Kanaya, H. Kawai, T. Kawai, M. Miyama, S. Miyashita,
M. Mori, Y. Nakamoto,
H. Nakamura, T. Nakamura,
I. Nakata, K. Nakazawa, K. Nemoto,
M. Okawa, A. Oshiyama,
Y. Oyanagi, S. Sakai,
T. Shirakawa,
A. Ukawa, M. Umemura, K. Wada,
Y. Watase, Y. Yamashita, M. Yasunaga,
and T. Yoshie.

\end{thebibliography}
\end{document}